\documentclass[twocolumn,showpacs,prl,floatfix]{revtex4}

\usepackage{graphicx}
\usepackage{bm}
\usepackage{psfrag}
\usepackage{amsmath}
\usepackage{amssymb}
\usepackage{latexsym}
\usepackage{exscale}

\newcommand{\vect}[1]{\bm{#1}}
\newcommand{\ten}[1]{\mbox{\textbf{
{\textsf{#1}}}}}

\newcommand{\sprod}{\!\cdot\!}
\newcommand{\tprod}{}
\newcommand{\vprod}{\!\times\!}
\newcommand{\trace}{\operatorname{tr}}
\newcommand{\trans}{\mathsf{T}}
\newcommand{\dif}{\mathrm{d}}
\newcommand{\mi}{\mathrm{i}}
\newcommand{\me}{\mathrm{e}}
\begin{document}

\title{Charge--Parity violating effects in Casimir--Polder potentials}

\author{Stefan Yoshi Buhmann}\email{stefan.buhmann@physik.uni-freiburg.de}
\affiliation{Physikalisches Institut, Albert-Ludwigs-Universit\"at
Freiburg, Hermann-Herder-Str. 3, 79104 Freiburg, Germany}
\affiliation{Freiburg Institute for Advanced Studies,
Albert-Ludwigs-Universit\"at Freiburg, Albertstr. 19, 79104
Freiburg, Germany}

\author{Valery N. Marachevsky}
\email{maraval@mail.ru} \affiliation{Department of Theoretical
Physics, Saint Petersburg State University, \\
Ulianovskaya 1, Petrodvorets, 198504 St.~Petersburg, Russia}

\author{Stefan Scheel}
\affiliation{Institut f\"ur Physik, Universit\"at Rostock,
Albert-Einstein-Stra{\ss}e 23, D-18059 Rostock, Germany}

\date{\today}

\begin{abstract}
We demonstrate under which conditions a violation of the charge--parity (CP)
symmetry in molecules will manifest itself in the Casimir--Polder interaction
of these with a magnetodielectric surface. Charge--parity violation induces a
specific electric--magnetic cross-polarisability in a molecule that is not
chiral, but time-reversal (T) symmetry violating. As we show, a detection of
such an effect via the Casimir--Polder potential requires a material medium
that is also sensitive to time-reversal, i.e., it must exhibit a non-reciprocal
electromagnetic response. As simple examples of such media we consider a
perfectly reflecting non-reciprocal mirror that is a special case of a perfect 
electromagnetic conductor as well as a Chern--Simons medium.
In addition, we show that Chern--Simons and related media can induce unusual 
atom--surface interactions for anisotropic molecules with and without a chiral 
response.
\end{abstract}


\maketitle


\paragraph{Introduction:}
The Casimir effect is an effective electromagnetic force between polarisable
objects that is induced by the recently directly observed vacuum fluctuations
of the quantum electromagnetic field \cite{Riek15}. Originally conceived by
Casimir as an attractive force between two perfectly conducting mirrors
\cite{Casimir48}, recent progress in material design and control has placed
Casimir forces between materials exhibiting both more realistic and more complex
electromagnetic responses in a focus of interest \cite{Woods15}. \emph{Inter
alia}, (para)magnetic media \cite{Revzen02,Joulain05}, chiral materials
\cite{Zhao09}, topological insulators \cite{Grushin11,Lu18} and graphene 
\cite{Bordag09,Fialkovsky11} have been studied.

A major driving force behind such investigations continues to be the search for
repulsive Casimir forces to overcome stiction in nanotechnology \cite{Serry98}.
Inspired by Boyer's observation that the force between a perfectly conducting
plate and an infinitely permeable one is repulsive \cite{Boyer69}, it was
theoretically predicted that repulsion persists for combinations of purely
electric and magnetic media with a more realistic response. Quite generally,
repulsive Casimir forces arise whenever the nature of the electromagnetic
reponse of the two interacting objects is diametrically opposite in a certain
sense, for example, if the objects show electric/magnetic responses, have
opposite chirality, or represent topological insulators with different internal
arrows of time.

Casimir--Polder forces between an atom or a molecule and a macroscopic body are
a closely related type of dispersion force \cite{CasimirPolder48} and hence
subject to the same phenomenology regarding their dependence on the
electromagnetic nature of the interacting objects. However, they are (i) a
local effect due to the small size of one of the two interacting objects, and
(ii) they can be measured with far higher accuracy due to the superior
techniques in manipulating and controlling single atoms or molecules
\cite{Fortagh07}. This suggests that Casimir--Polder forces might (i) serve as
a probe of the molecule's or the body's properties rather than a force that is
to be manipulated and overcome, and (ii) that this probe could be applied to
access very exotic properties of matter. In particular, we intend to apply this
idea to the phenomenon of charge-parity violation in molecules, asking the
hypothetical question: Would a potential charge-parity violation in a molecule
become manifest in its Casimir--Polder interaction with a macroscopic body and,
if so, under what conditions?

One of the fundamental symmetries of the standard model of particle
physics is the invariance under combined charge (C), parity (P) and
time (T) reversal. However, a physical system need not be invariant
under each of these three symmetries individually. Examples of
parity non-conservation include processes involving the weak force
such as the $\beta$ decay \cite{Lee56,Wu57}. The weak interaction is
also responsible for the broken parity invariance in rovibrational
spectra of chiral molecules \cite{Quack00,Darquie10}. The
antiferromagnet $\mathrm{Cr_2O_3}$, on the other hand, is an example
of a P odd and T odd system that exhibits a pseudoscalar response
(see Ref.~\cite{Hehl08} and references therein) which provides a
template for perfect electromagnetic conductors \cite{Sihvola08}. In
the search for physics beyond the standard model, weak-scale
supersymmetry would induce a CP violation that is reflected in an
intrinsic electron dipole moment (EDM) of neutrons or electrons (see
Ref.~\cite{Pospelov05} for a review). Recent experiments place an
upper limit of $|d_e|<8.7\times10^{-29}e$cm on the electron EDM
\cite{EDM2}.

Within a field-theoretic setting, it is known that a Chern--Simons
interaction violates both P and T symmetries
\cite{Khriplovich1,Khriplovich2}. The Casimir interaction
between two flat Chern--Simons layers was studied in
Refs.~\cite{MarkovPismak, Mar7, Mar8, DV}, with its main
feature being the prediction of both attractive and repulsive
regimes of the Casimir force between Chern--Simons layers separated
by a vacuum slit. A physical example showing an effective
Chern--Simons interaction is a quantum Hall system consisting of a
2D electron gas \cite{Jackiw}. The coupling constant of
the Chern--Simons action is different for each quantum Hall plateau,
with its value being determined by the external magnetic field
perpendicular to the quantum Hall layer. The implications is that
the Casimir-Polder potential will be quantized at the quantum Hall
plateaus \cite{Cysne14}.

In this article we investigate the Casimir--Polder potential for an anisotropic
molecule in the presence of a flat Chern--Simons layer. It is known that the
CP-conserving part of the Casimir-Polder potential is quadratic in the coupling
constant of the Chern--Simons layer \cite{Valery}. On the other hand, the
CP-violating part of the Casimir--Polder potential, that we derive here, is
linear in the coupling constant. The sign of the Chern--Simons coupling can be
altered by reversing the direction of an external magnetic field. As a result,
the CP-violating part of the Casimir-Polder potential can be extracted from
measurements at any plateau of a quantum Hall system performed at external
magnetic fields with alternating spatial directions.

The article is organised as follows. We begin with a discussion
on atom--field coupling and the resulting Casimir--Polder interaction
potentials in the presence of non-reciprocal media. As a particular example, we
consider a planar Chern--Simons layer that gives rise to non-reciprocal
effects. In the following, we then construct the Casimir--Polder potentials for
molecules with various anisotropic, asymmetric polarisabilities, including the
particular case of CP-violating molecules. We close the article with some
concluding remarks. Details regarding the Chern--Simons action and its influence on Maxwell's 
equations as well as the definition of CP-violating polarisabilities have been 
delegated into the Supplementary Online Material (SOM).

\paragraph{Atom--field coupling:}
The Curie dissymmetry principle \cite{1006} suggests that CP-violating atomic
properties can only couple to environments which are also T violating, i.e.,
non-reciprocal, so that the Green's tensor does not necessarily fulfil the
Onsager relation $\ten{G}^\trans(\vect{r}',\vect{r},\omega)
\!=\!\ten{G}(\vect{r},\vect{r}',\omega)$ regarding the reversibility of optical
paths \cite{Onsager}. This suggests a novel possibility for the detection of CP-violating
atomic properties by studying the electromagnetic interaction of a CP-violating
molecule with a macroscopic Chern--Simons layer by means of the Casimir--Polder
potential.

The interaction of a molecule $A$ with the electromagnetic field can be
described by the multipolar Hamiltonian in long-wavelength approximation as
\cite{Scheel08}
\begin{equation}
\label{c1}
\hat{H}_{AF}=
-\hat{\vect{d}}\sprod\hat{\vect{E}}(\vect{r}_{\!A})
-\hat{\vect{m}}\sprod\hat{\vect{B}}(\vect{r}_{\!A})
\end{equation}
($\hat{\vect{d}}$, $\hat{\vect{m}}$: molecular electric and magnetic
dipole moments; $\vect{r}_A$: position) when neglecting the diamagnetic
interaction. The position-dependent Casimir--Polder potential can be derived
within second-order perturbation theory by replacing the full propagator of the
electromagnetic field $\ten{G}$ with its scattering part $\ten{G}^{(1)}$
\cite{Scheel08,Valery2}. In the following, we will derive contributions to
the Casimir--Polder potential from electric-electric and electric-magnetic
terms in the presence of a planar Chern--Simons layer.


\paragraph{Casimir--Polder potential for a non-magnetic molecule:}
The Casimir--Polder potential \cite{CasimirPolder48} for a non-magnetic ground-state
molecule arises from the second-order energy shift
\begin{equation}
\label{cp11}
U_{ee}(\vect{r}_{\!A})=\frac{\hbar\mu_0}{2\pi}
 \int_0^\infty\dif\xi\,\xi^2 \trace\bigl[\bm{\alpha}(\mi\xi)
 \sprod\ten{G}^{(1)}(\vect{r}_{\!A},\vect{r}_{\!A},\mi\xi)\bigr]
\end{equation}
where
\begin{equation}
\label{cp12}
\bm{\alpha}(\omega)
=\lim_{\epsilon\to 0+}\frac{1}{\hbar}\sum_k\biggl[
 \frac{\vect{d}_{k0}\tprod\vect{d}_{0k}}
 {\omega+\omega_k+\mi\epsilon}
 -\frac{\vect{d}_{0k}\tprod\vect{d}_{k0}}
 {\omega-\omega_k+\mi\epsilon}
 \biggr]
\end{equation}
is the molecular polarisability, and $\ten{G}^{(1)}$ is the scattering
part of the electromagnetic Green's tensor. In order to
make the influence of non-reciprocal media more explicit, we decompose the
polarisability and Green's tensors into their respective symmetric and
antisymmetric parts:
\begin{multline}
\label{cp13}
U_{ee}(\vect{r}_{\!A})=\frac{\hbar\mu_0}{2\pi}
 \int_0^\infty\dif\xi\,\xi^2 \trace\bigl[\bm{\alpha}_S(\mi\xi)
 \sprod\ten{G}^{(1)}_S(\vect{r}_{\!A},\vect{r}_{\!A},\mi\xi)\\
 +\bm{\alpha}_A(\mi\xi)
 \sprod\ten{G}^{(1)}_A(\vect{r}_{\!A},\vect{r}_{\!A},\mi\xi)\bigr]\;.
\end{multline}
The first term is the ordinary Casimir--Polder potential in environments
respecting the Onsager theorem \cite{0042,0046}. The second term is due to
the presence of non-reciprocal media; it only arises for molecules with an
anisotropic, asymmetric polarisability. Examples for its relevance are the
recently considered interaction of an atom with a plate exhibiting a
Chern--Simons interaction \cite{Valery} or with a topological-insulator plate 
\cite{Fuchs17b,Ruiz18}.

The scattering part $\ten{G}^{(1)}$ in the presence of a Chern--Simons plane layer is given 
in the SOM. The symmetric part of the polarisability
leads in this case to a potential equal to the Casimir-Polder potential in
front of a perfectly conducting plate multiplied by a factor $a^2/(1+a^2)$. The
asymmetric part of the polarisability yields an additional interaction with the
Chern--Simons plate \cite{Valery}, viz.
\begin{multline}
U_{as}(z_A) = \frac{\hbar}{32 \pi^2 \varepsilon_0 c}\frac{a}{1+a^2} \\
\times\int_0^{+\infty} d\xi \epsilon_{jlz} \alpha_{jl}(i\xi) \xi \Bigl(1+2
\frac{\xi z_A}{c}\Bigr) \me^{-2 \xi z_A/c} . \label{P1}
\end{multline}

In the retarded limit, $\omega_k z_A/c \gg 1$, the potential
(\ref{P1}) leads to an $1/z_A^5$ asymptote
\begin{equation} \label{P2}
U_{as}(z_A) = - \frac{c^2}{8\pi^2 \varepsilon_0 z_A^5}\frac{a}{1+a^2}
\sum_k \frac{{\rm Im}( d_{0k,x} d_{k0,y})
}{\omega_k^2},
\end{equation}
whereas at short separations, $\omega_k z_A/c \ll 1$, it is well approximated
by a $1/z_A^3$ potential
\begin{equation} \label{P3}
U_{as}(z_A) = -\frac{1}{16\pi^2 \varepsilon_0
z_A^3}\frac{a}{1+a^2}\sum_k {\rm Im}(d_{0k,x}d_{k0,y}) .
\end{equation}
The limiting cases of a perfectly reflecting, non-reciprocal mirror
($r_{s \to p}=\pm 1, r_{p \to s}=\pm 1, r_s=0, r_p=0$) can be
immediately obtained from Eqs.~(\ref{P1})---(\ref{P3}) by substituting
$a/(1+a^2) \to \pm 1$.
The latter is a specific example of a perfect electromagnetic conductor 
\cite{Sihvola08} and emerges from a perfect electric conductor by means of a 
duality transformation \cite{Sihvola08}.


\paragraph{Casimir--Polder potential for a molecule with
CP-violating cross-polarisabilities:}
For an electromagnetic molecule, both electric and magnetic dipole
couplings contribute to the atom--field interaction (\ref{c1}). Beyond the
purely electric interaction, we are now interested in the part of the
Casimir--Polder potential due to the second-order energy shift arising from
mixed electric--magnetic transitions:
\begin{multline}
\label{cp16}
U_{CP}(\vect{r}_{\!A})=U_{em}(\vect{r}_{\!A})+U_{me}(\vect{r}_{\!A})\\
 =-\frac{\hbar\mu_0}{2\pi}\int_0^\infty\!\!\!\dif\xi\,\xi
 \Bigl\{\trace\bigl[\bm{\chi}_{me}(\mi\xi)\sprod
 \ten{G}^{(1)}(\vect{r}_{\!A},\vect{r}_{\!A},\mi\xi)
 \vprod\overleftarrow{\vect{\nabla}}'\bigr]\\
+\trace\bigl[\bm{\chi}_{em}(\mi\xi)\sprod\vect{\nabla}\vprod
 \ten{G}^{(1)}(\vect{r}_{\!A},\vect{r}_{\!A},\mi\xi)\bigr]\Bigr\}
\end{multline}
where we have introduced the cross-polarisabilities
\begin{gather}
\label{cp17}
\bm{\chi}_{em}(\omega)
=\lim_{\epsilon\to 0}\frac{1}{\hbar}\sum_k
 \biggl[\frac{\vect{d}_{k0}\tprod\vect{m}_{0k}}
 {\omega+\omega_k+\mi\epsilon}
 -\frac{\vect{d}_{0k}\tprod\vect{m}_{k0}}
 {\omega-\omega_k+\mi\epsilon}\biggr],\\
\label{cp18}
\bm{\chi}_{me}(\omega)
=\lim_{\epsilon\to 0}\frac{1}{\hbar}\sum_k
 \biggl[\frac{\vect{m}_{k0}\tprod\vect{d}_{0k}}
 {\omega+\omega_k+\mi\epsilon}
 -\frac{\vect{m}_{0k}\tprod\vect{d}_{k0}}
 {\omega-\omega_k+\mi\epsilon}\biggr]\,.
\end{gather}

It is customary to decompose the Hamiltonian
$\hat{H}_A=\hat{H}_0+\hat{V}^\mathrm{CP}$ of the atomic subsystem
into CP-conserving and violating parts \cite{Derevianko10}. The
cross-polarisabilities can then be expressed in terms of the eigenstates
$|n^0\rangle$ and energies $E_n^0$ of the CP-conserving Hamiltonian
(see SOM):
\begin{multline}
\label{cp17b}
\bm{\chi}_{em}(\omega)\\
=\lim_{\epsilon\to 0}\frac{1}{\hbar}\sum_{k,l}
 \biggl[\frac{V^\mathrm{CP}_{0l}\vect{d}^0_{kl}\tprod\vect{m}^0_{0k}}
 {\omega^0_l(\omega+\omega^0_k+\mi\epsilon)}
 -\frac{V^\mathrm{CP}_{0l}\vect{d}^0_{lk}\tprod\vect{m}^0_{k0}}
 {\omega^0_l(\omega-\omega^0_k+\mi\epsilon)}\biggr]\\
+\lim_{\epsilon\to 0}\frac{1}{\hbar}\sum_{k,l}
 \biggl[\frac{\vect{d}^0_{l0}\tprod V^\mathrm{CP}_{lk}
 \vect{m}^0_{0k}}
 {(\omega+\omega^0_l+\mi\epsilon)(\omega+\omega^0_k+\mi\epsilon)}\\
 -\frac{\vect{d}^0_{0l}\tprod V^\mathrm{CP}_{lk}\vect{m}^0_{k0}}
 {(\omega-\omega^0_l+\mi\epsilon)(\omega-\omega^0_k+\mi\epsilon)}
 \biggr]\\
+\lim_{\epsilon\to 0}\frac{1}{\hbar}\sum_{k,l}
 \biggl[\frac{\vect{d}^0_{k0}\tprod\vect{m}^0_{lk}V^\mathrm{CP}_{l0}}
 {\omega^0_l(\omega+\omega^0_k+\mi\epsilon)}
 -\frac{\vect{d}^0_{0k}\tprod\vect{m}^0_{kl}V^\mathrm{CP}_{l0}}
 {\omega^0_l(\omega-\omega^0_k+\mi\epsilon)}\biggr]
\end{multline}
with the definitions $\omega^0_k=(E^0_k-E^0_0)/\hbar$,
$\vect{d}^0_{nm}=\langle n^0|\hat{\vect{d}}|m^0\rangle$,
$\vect{m}^0_{nm}=\langle n^0|\hat{\vect{m}}|m^0\rangle$
and $V^\mathrm{CP}_{nm}
=\langle n^0|\hat{V}^\mathrm{CP}|m^0\rangle$.

For a CP-violating system, the electric and magnetic transition dipole matrix
elements have a vanishing relative phase, so that
$\bm{\chi}_{me}=\bm{\chi}_{em}^\trans$. Due to this symmetry, it is clear that
the cross-polarisabilities of a CP-violating system do not lead to an
interaction with a perfectly conducting plate. However, they do provide an
interaction with a Chern--Simons layer as well as with a perfectly reflecting
non-reciprocal mirror. To evaluate this contribution of the 
cross-polarisabilities of a CP-violating molecule to the Casimir--Polder
interaction with a non-reciprocal medium, we use Eq.~(\ref{cp16}) with  
$\chi_{jl} \equiv \chi_{em,jl} = \chi_{me,lj}$ to find
\begin{multline} \label{cp23}
U_{CP}(z_A) = \frac{\hbar}{32 \pi^2 \varepsilon_0 c z_A^3}
\frac{a}{1+a^2}\int_0^{\infty} d\xi \me^{-2 \xi z_A/c} \\ \times
 \biggl\{
\bigl[\chi_{xx}(i\xi)+ \chi_{yy}(i\xi)\bigr]
 \biggl(1+2\frac{\xi z_A}{c} + 4 \frac{\xi^2
z_A^2}{c^2}\biggr)  \\+ 2 \chi_{zz}(i\xi)\biggl(1+2 \frac{\xi z_A}{c}\biggr) \biggr\}.
\end{multline}
Note that the symmetry of the cross-polarisabilities implies that
$\chi_{jl}(i\xi) = \frac{1}{\hbar}\sum_k
\frac{2\omega_k}{\omega_k^2 + \xi^2} d_{k0,j} m_{0k,l}$ where
$d_{k0,j}m_{0k,l}$ is a real number. In the retarded limit,
$\omega_kz_A/c\gg 1$, the approximation
$\chi_{jl}(\mi\xi)\simeq\chi_{jl}(0)$ leads to the asymptote
\begin{multline}
\label{cp24} U_{CP}(z_A)
=\frac{\hbar}{16\pi^2\varepsilon_0z_A^4} \frac{a}{1+a^2} \\
\times \bigl[\chi_{xx}(0)+\chi_{yy}(0)+\chi_{zz}(0)\bigr].
\end{multline}
In the opposite nonretarded limit, we may approximate
\begin{multline}
\label{cp25} U_{CP}(z_A) \!=\!\frac{\hbar}{32\pi^2\varepsilon_0
cz_A^3} \frac{a}{1+a^2} \\ \times
 \int_0^\infty\dif\xi\,\bigl[\chi_{xx}(\mi\xi) + \chi_{yy}(\mi\xi) +
 2\chi_{zz}(\mi\xi)\bigr] .
\end{multline}
The limiting case of a perfectly reflecting, non-reciprocal mirror ($r_{s \to
p}=\pm 1, r_{p \to s}=\pm 1, r_s=0, r_p=0$) can be immediately
obtained from Eqs.~(\ref{cp23})--(\ref{cp25}) by replacing
$a/(1+a^2) \to \pm 1$.

Upon substitution $ \bm{\chi} \to \bm{\alpha}$, the potential
(\ref{cp23}) coincides with the well known Casimir--Polder potential
\cite{CasimirPolder48} of a purely electric atom in front of a perfectly conducting plate, 
apart from a factor two. This correspondence can easily be understood from the 
duality of electric and magnetic fields \cite{Onsager}. Under a duality 
transformation by an angle $\theta/4$, a perfectly conducting plate transforms 
to a perfect nonreciprocal reflector \cite{Sihvola08} while a purely dielectric 
atom transforms into one with cross-polarisabilities. The factor two stems from 
the fact that two cross-polarisabilities contribute to $U_{CP}$ as opposed to 
the single electric polarisability contributing to the ordinary Casimir--Polder 
potential.

\paragraph{Chern-Simons interaction with chiral molecules:}
For chiral, time-reversal invariant molecules, Lloyd's theorem states that
electric and magnetic transitions carry a relative phase factor
$\mi=\me^{\mi\pi/2}$ \cite{0867}, so the relation
$\bm{\chi}_{me}=-\bm{\chi}_{em}^\trans$ holds between the
cross-polarisabilities. The case of isotropic chiral polarisability and
respective chiral Casimir--Polder potential was studied in
Ref.~\cite{chiral}. Here we demonstrate a novel possibility. It turns out that
the interaction of a molecule with an anisotropic, asymmetric chiral
polarisability and non-chiral media leads to an additional component of the
Casimir-Polder potential. As an example, we evaluate this new component of the
Casimir-Polder potential for a molecule with anisotropic, asymmetric chiral
polarisability in front of a Chern--Simons layer.

Using Eq.~(\ref{cp16}) one obtains the Chern--Simons interaction with
P-violating chiral molecules now satisfying $\chi_{jl} \equiv
\chi_{em,jl} = - \chi_{me,lj}$ :
\begin{multline} \label{W5}
U_{P}(z_A) =
-\frac{\hbar}{64 \pi^2 \varepsilon_0 z_A^4} \frac{a^2}{1+a^2}
\int_0^{+\infty} d\xi \epsilon_{jlz} \chi_{jl}(\mi\xi)
\\ \times \frac{e^{- 2 \xi z_A/c}}{\xi}  \biggl(3 + 6 \frac{\xi z_A}{c}+
8\frac{\xi^2 z_A^2}{c}+ 8 \frac{\xi^3 z_A^3}{c} \biggr)
\end{multline}
where a summation over $j,l$ is implied. Note that due to the symmetry
$\chi_{em,jl} = - \chi_{me,lj}$, one can write
$\chi_{jl}(i\xi) = \frac{1}{\hbar}\sum_k \frac{2\xi}{\omega_k^2 +
\xi^2} {\rm Im}(d_{k0,j}m_{0k,l})$, where
$d_{k0,j}m_{0k,l}$ is purely imaginary. As a result, in
the retarded limit, $\omega_k z_A/c\gg 1$, we obtain
\begin{equation}
U_{P}(z_A) = - \frac{c}{4\pi^2 \varepsilon_0 z_A^5}
\frac{a^2}{1+a^2}
\sum_k \frac{\epsilon_{jlz}{\rm
Im}(d_{k0,j}m_{0k,l})}{\omega_k^2}  .
\end{equation}
In the opposite nonretarded limit we may approximate
\begin{multline}
U_{P}(z_A) = -\frac{3 \hbar}{64\pi^2\varepsilon_0 z_A^4}
\frac{a^2}{1+a^2} \int_0^{+\infty} d\xi
\frac{\epsilon_{jlz}\chi_{jl}(\mi\xi)
}{\xi}\\
=-\frac{3}{64\pi\varepsilon_0 z_A^4} \frac{a^2}{1+a^2} \sum_k
\frac{\epsilon_{jlz}{\rm Im}(d_{k0,j}m_{0k,l})}{\omega_k}  .
\end{multline}
In the limit $a \to \pm\infty$ we obtain the potential of a chiral
molecule in front of a perfectly conducting plate. Note, however, that
the quantity $\chi_{xy}(\mi\xi) - \chi_{yx}(\mi\xi)$ should be different from zero
to obtain a non-vanishing potential.

\paragraph{Conclusions:}
With regard to answering our central question, we have shown that charge-parity 
violating effects in molecules can indeed be manifest in their Casimir--Polder 
interaction with a surface. As anticipated from the Curie dissymmetry 
principle, this requires the surface to also possess CP or T-violating 
properties. We  have shown this explicitly for a nonreciprocal perfect 
reflector as an example of a perfect electromagnetic conductor medium
as well as for a Chern--Simons medium. The result in the former case is 
strikingly similar in form to the well-known formula by Casimir and Polder,
which can readily be understood from the duality invariance of QED in the 
absence of free charges or currents.

In addition, we have shown that Chern--Simons media lead to a new
and previously unobserved Casimir--Polder potential for anisotropic
chiral molecules. In this case, the respective power laws of the
potential in the short and long-distance limits differ from those
previously predicted for isotropic molecules. Our findings can be
generalised to topological insulator media, where we expect similar
new potential components for CP-violating or anisotropic molecules.

The CP-violating polarisability that induces the respective
potential is estimated to be very small \cite{Flambaum}. In order to
use the Casimir--Polder interaction as a probe for CP violation, one
would have to enhance the effect and the sensitivity by measuring
the Casimir--Polder-induced frequency shift spectroscopically in
a resonating cavity.

\acknowledgments We would like to thank V.~Flambaum, O.P.~Sushkov, A.N.~Petrov,
and L.V.~Skripnikov for discussions. S.Y.B is
grateful by support by the German Research Foundation (DFG, Grant
No. BU 1803/3-1) and by the Freiburg Institute of Advanced Studies.
V.N.M. carried out research using computational resources provided
by the Computer Center of SPbU (http://cc.spbu.ru/en) through a grant of Saint 
Petersburg State University No.$11.40.538.2017$.

%
%

\end{document}


\title{Charge--Parity violating effects in Casimir--Polder potentials 
(Supplementary Online Material)}

\author{Stefan Yoshi Buhmann}\email{stefan.buhmann@physik.uni-freiburg.de}
\affiliation{Physikalisches Institut, Albert-Ludwigs-Universit\"at
Freiburg, Hermann-Herder-Str. 3, 79104 Freiburg, Germany}
\affiliation{Freiburg Institute for Advanced Studies,
Albert-Ludwigs-Universit\"at Freiburg, Albertstr. 19, 79104
Freiburg, Germany}

\author{Valery N. Marachevsky}
\email{maraval@mail.ru} \affiliation{Department of Theoretical
Physics, Saint Petersburg State University, \\
Ulianovskaya 1, Petrodvorets, 198504 St.~Petersburg, Russia}

\author{Stefan Scheel}
\affiliation{Institut f\"ur Physik, Universit\"at Rostock,
Albert-Einstein-Stra{\ss}e 23, D-18059 Rostock, Germany}

\date{\today}

\maketitle


\section{Diffraction from a Chern-Simons layer}
The action for a planar Chern--Simons layer at $z=0$ has the form
\begin{equation}
S=\frac{a}{2} \int \varepsilon^{z\nu\rho\sigma} A_\nu F_{\rho\sigma}
\dif t \dif x \dif y, \label{CS1}
\end{equation}
where the current due to the Chern--Simons interaction is
$J^\nu= a \varepsilon^{z \nu \rho \sigma} F_{\rho\sigma}$. Maxwell's equations
for the electromagnetic field in the presence of the action (\ref{CS1}) are
subsequently modified to
\begin{equation}
\partial_\mu F^{\mu \nu} + a\, \varepsilon^{z \nu \rho\sigma}
F_{\rho\sigma} \delta(z) =0. \label{motion}
\end{equation}
From Eq.~(\ref{motion}), the continuity conditions follow as (see
e.g. Ref.~\cite{BC}):
\begin{align}
E_z|_{z=0^+} - E_z|_{z=0^-} &= - 2 a H_z|_{z=0}, \label{cont1}\\
H_x|_{z=0^+} - H_x|_{z=0^-} &=  2 a E_x|_{z=0}, \label{cont2}\\
H_y|_{z=0^+} - H_y|_{z=0^-} &=  2 a E_y|_{z=0}. \label{cont3}
\end{align}

Let us consider an $s$-polarised plane electromagnetic wave impinging onto a
planar Chern--Simons layer at $z=0$ (the factor $\exp(i(\omega/c) t + i k_y y)$
is omitted for simplicity):
\begin{align}
E_x &= \exp(-\mi k_z z) + r_{s} \exp(i k_z z) , z>0,\label{wave1}  \\
E_x &= t_{s} \exp(-\mi k_z z) ,  z<0,   \label{wave2} \\
H_x &= r_{s \to p} \exp(\mi k_z z) , z>0, \label{wave3}  \\
H_x &= t_{s \to p} \exp(-\mi k_z z) , z<0, \label{wave4}
\end{align}
By applying the continuity conditions (\ref{cont1})--(\ref{cont3})
one obtains the reflection and transmission coefficients of an $s$-polarised
wave as
\begin{align}
&r_{s} = - \frac{a^2}{1+a^2}, &  &t_{s} = \frac{1}{1+a^2} ,
\nonumber\\
&r_{s \to p} = \frac{a}{1+a^2}, &  &t_{s \to p} = - \frac{a}{1+a^2}.
\end{align}
By a duality transformation, i.e. by exchanging the fields $E$, $H$ as well as
the indices $s$, $p$ in Eqs.~(\ref{wave1})-(\ref{wave4}), we obtain the reflection and transmission
coefficients the diffraction of a $p$-polarised wave as
\begin{align}
&r_{p} =  \frac{a^2}{1+a^2}, &  &t_{p} = \frac{1}{1+a^2} ,
\nonumber\\
&r_{p \to s} = \frac{a}{1+a^2}, &  &t_{p \to s} = \frac{a}{1+a^2}.
\end{align}

\section{Scattering Green function for a planar Chern--Simons layer}

 The scattering Green function above ($z_A>0$) a Chern--Simons plate in the
gauge $A_0=0$  mixes $s$- and $p$- polarised waves:
\begin{multline}
\label{cp21}
\ten{G}^{(1)}(\vect{r},\vect{r}',\mi\xi)
=\frac{1}{8\pi^2}\int\frac{\dif^2q}{\beta}\,
 \me^{\mi\vect{q}\cdot(\vect{r}-\vect{r}')-\beta(z+z')}\\
\times(\vect{e}_s^+\vect{e}_s^-r_{s} + \vect{e}_p^+\vect{e}_p^-r_{p}
 +\vect{e}_p^+\vect{e}_s^-r_{s\to
p} +\vect{e}_s^+\vect{e}_p^-r_{p\to s})
\end{multline}
($\vect{q}\perp\vect{e}_z$, $\beta=\sqrt{\xi^2/c^2+q^2}$) where the
polarisation unit vectors for $s$- and $p$-polarised waves read
$\vect{e}_s^\pm=\vect{e}_q\vprod\vect{e}_z$,
$\vect{e}_p^\pm=-(c/\xi)(\mi q\vect{e}_z\pm\beta\vect{e}_q)$] and
the respective reflection coefficients of the half space are given
above.
A general technique for calculating the Casimir-Polder potential in
an arbitrary gauge of the vector potentials was developed in
Ref.~\cite{Valery}, with example calculations in different gauges being
provided in Ref.~\cite{Valery2}.

\section{CP-violating polarisability}
Upon decomposing $\hat{H}_A=\hat{H}_0+\hat{V}^\mathrm{CP}$, we can
use perturbation theory to express the eigenstates $|n\rangle$ and
energies $E_n$ of $\hat{H}_A$ in terms of the eigenstates
$|n^0\rangle$ and energies $E_n^0$ of $\hat{H}_0$. Assuming that
$\langle n|\hat{V}^\mathrm{CP}|n\rangle\equiv V^\mathrm{CP}_{nn}=0$
\cite{Derevianko10}, we have $E_n=E_n^0$ in linear order in
$\hat{V}^\mathrm{CP}$ while the eigenstates acquire linear shifts due
to the CP-violating interaction:
\begin{equation}
|n\rangle=|n^0\rangle
 +\sum_l|l^0\rangle\frac{\langle l^0|\hat{V}^\mathrm{CP}|n^0\rangle}
{E^0_n-E^0_l}\,.
\end{equation}
To linear order in $\hat{V}^\mathrm{CP}$, we can hence expand
\begin{multline}
\label{cp17c}
\vect{d}_{0k}\tprod\vect{m}_{k0}
=-\sum_{l}
 \frac{V^\mathrm{CP}_{0l}\vect{d}^0_{kl}\tprod\vect{m}^0_{0k}}
 {\hbar\omega^0_l}
-\sum_{l}
 \frac{\vect{d}^0_{0l}\tprod V^\mathrm{CP}_{lk}\vect{m}^0_{k0}}
 {\hbar(\omega^0_l-\omega^0_k)}\\
-\sum_{l}
 \frac{\vect{d}^0_{0k}\tprod V^\mathrm{CP}_{kl}\vect{m}^0_{l0}}
 {\hbar(\omega^0_l-\omega^0_k)}
-\sum_{l}
 \frac{\vect{d}^0_{0k}\tprod\vect{m}^0_{kl}V^\mathrm{CP}_{l0}}
 {\hbar\omega^0_l}\,.
\end{multline}
Substituting this result into Eq.~(9) in the main article,
relabelling $l\leftrightarrow k$ in the third term, and using the
identity
\begin{multline}
\frac{1}{(\omega^0_l-\omega^0_k)(\omega^0_k\pm\omega+\mi\epsilon)}
+\frac{1}{(\omega^0_l-\omega^0_k)(\omega^0_l\pm\omega+\mi\epsilon)}\\
=\frac{1}{(\omega^0_k\pm\omega+\mi\epsilon)
 (\omega^0_l\pm\omega+\mi\epsilon)}
\end{multline}
results in Eq.~(11) in the main article.